\documentclass[letterpaper, conference, 10 pt]{ieeeconf}
\IEEEoverridecommandlockouts
\usepackage[english]{babel}

\usepackage{amsmath, amssymb, amsthm}
\usepackage{cite}
\usepackage{algorithmic}
\usepackage{graphicx}
\usepackage{caption}
\usepackage{subcaption}
\usepackage{textcomp}
\usepackage{xcolor}
\let\labelindent\relax

\usepackage{enumitem}

\def\BibTeX{{\rm B\kern-.05em{\sc i\kern-.025em b}\kern-.08em
		T\kern-.1667em\lower.7ex\hbox{E}\kern-.125emX}}

\usepackage{optidef}
\def\matt#1{\begin{bmatrix}#1\end{bmatrix}}

\newtheorem{defn}{Definition}
\newtheorem{rem}[defn]{Remark}
\newtheorem{lem}[defn]{Lemma}

\newtheorem{assum}[defn]{Assumption}

\newtheorem{thm}[defn]{Theorem}

\title{\LARGE \bf
	Cooperative Tuning of Multi-Agent Optimal Control Systems
}

\author{Zehui Lu$^{1}$, \ Wanxin Jin$^{2}$, \ Shaoshuai Mou$^{1}$, \ Brian D. O. Anderson$^{3}$
	\thanks{$^{1}$Zehui Lu and Shaoshuai Mou are with the School of Aeronautics and Astronautics, Purdue University, IN 47907, USA {\tt\small \{lu846, mous\}@purdue.edu}}
	\thanks{$^{2}$Wanxin Jin is with the GRASP Laboratory, University of Pennsylvania, Philadelphia, PA 19104, USA {\tt\small wanxinjin@gmail.com}}
	\thanks{$^{3}$Brian D. O. Anderson is with The Australian National University, Acton, ACT 2601, Australia {\tt\small brian.anderson@anu.edu.au}}
	\thanks{This work was supported in part by grants from the NASA University Leadership Initiative (ULI), Northrop Grumman Corporation, and Rolls-Royce Corporation.}
}

\begin{document}
	
	\maketitle
	\thispagestyle{empty}
	\pagestyle{empty}

	\begin{abstract}
		This paper investigates the problem of cooperative tuning of multi-agent optimal control systems, where a network of agents (i.e. multiple coupled optimal control systems) adjusts parameters in their dynamics, objective functions, or controllers in a coordinated way to minimize the sum of their loss functions. Different from classical techniques for tuning parameters in a controller, we allow tunable parameters appearing in both the system dynamics and the objective functions of each agent. A framework is developed to allow all agents to reach a consensus on the tunable parameter, which minimizes team loss. The key idea of the proposed algorithm rests on the integration of consensus-based distributed optimization for a multi-agent system and a gradient generator capturing the optimal performance as a function of the parameter in the feedback loop tuning the parameter for each agent. Both theoretical results and simulations for a synchronous multi-agent rendezvous problem are provided to validate the proposed method for cooperative tuning of multi-agent optimal control.
	\end{abstract}

\section{Introduction} \label{section:introduction}
\emph{Optimal control} theories are developed to find control inputs for a plant such that its states and inputs optimize a particular objective\cite{anderson1990optimal}. An optimal control (OC) system typically includes system dynamics and an objective function to be optimized given a task specification. The system dynamics and objective function can determine an optimal controller for a given OC system.  In order to employ a pre-designed OC system in practice, one often allows tunable parameters appearing in one or more of its dynamic model, objective function, or the optimal controller (if it is available) to tune the OC system to meet additional performance requirements or even to minimize an additional performance index.
\emph{Tuning} of OC systems refers to the adjustment of a tunable parameter in the control system to further minimize such an additional performance index while retaining optimality (through some adjustment of the optimal control for the original performance index to reflect the tunable parameter adjustment).
Such an additional performance index commonly involves a scalar \emph{loss function} defined in terms of a system's instantaneous states/inputs to serve as a criterion to evaluate the system's performance, and could reflect stability, fast and smooth set-point tracking, robustness for disturbance rejection, mission changes or newly arisen safety constraints. Tuning OC systems is critical in adapting OC to different application scenarios and the hope is that it can be achieved without re-designing the OC system from the beginning. 

Tuning of OC system is in the tradition of what has become known as neighboring extremal optimal control (NEOC). Reference \cite{bryson2018applied} (the first edition of which originally appeared in 1975) treats the following problem. Suppose an open-loop optimal control is known for a nonlinear system with prescribed initial condition, and suppose the initial condition is then varied by a small amount; how can one obtain (easily) a corresponding small variation to the control to maintain optimality? The answer rests on what is known as the theory of the second variation, and boils down to solving a time-varying linear-quadratic optimal control problem with parameters derived from the original problem and its optimum trajectory. From this consideration of perturbations of the initial condition, attention moved to perturbations of other aspects of optimal control problems, including parameters in the loss function. Thus reference \cite{rehbock1992computational} considers a nonlinear optimal control problem in which there are one or more scalar parameters which potentially can vary. With a solution available for one set of parameter values, an algorithm is given whereby the gradient of the optimal index with respect to those parameters is computed. It is a variant on that provided in \cite{bryson2018applied} for initial condition perturbation, on a time-varying linear-quadratic foundation.  In a further example,  \cite{fisher1995neighbouring}  indicates how NEOC can work in the presence of control constraints. 

Building on the early focus on small variations in initial conditions or parameters, the paper \cite{jiang2015optimal} considers a nonlinear optimal control problem where a parameter undergoes a significant change from the value used to compute the optimal control. It is shown how a modified optimal control can be computed using multiple applications of the neighboring extremal method, each corresponding to a distinct point on a homotopic path, and also importantly demonstrates the possibility that a change of extremal due to an infinitesimal change in the parameter can be  discontinuous, though the performance index value may be continuous across the change. 

By and large, these papers all work with continuous time systems, but unsurprisingly the ideas carry over to discrete time \cite{ghaemi2009neighboring}. Very recently the authors of \cite{jin2020pontryagin} have developed a framework for tuning of an OC system based on differentiating the Pontryagin's Maximum Principle corresponding to the OC system. Different from classical research in tuning parameters in controllers \cite{kazantzis2005optimal}, objective functions (known as learning from demonstrations  \cite{jin2019inverse,jin2021inverse,jin2021distributed} ), or system dynamics (known as system identification \cite{schon2011system,  abraham2019active}), the work in \cite{jin2020pontryagin} allows tunable parameters existing in controllers, objective functions and system dynamics. In this paper we aim to further extend the result in \cite{jin2020pontryagin} from tuning of an OC system to cooperative tuning of multiple coupled multi-agent OC systems.  

\begin{figure}[th]
	\centering
	\includegraphics[width=0.4\textwidth]{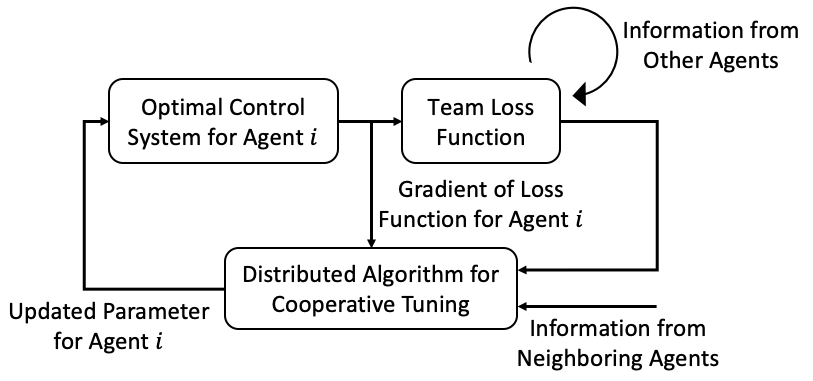}
	\caption{Cooperative Tuning of Multi-Agent Optimal Control Systems (CT-MAOCS)}
	\label{fig/multi_agent_problem}
\end{figure}

By working as a cohesive whole, a \emph{multi-agent system} can usually accomplish complicated missions well beyond capabilities of individual subsystems \cite{mou2015distributed,wang2019scalable}. But there is little work on the problem of \emph{cooperative tuning of multi-agent optimal control systems (CT-MAOCS)}. The scenario to be considered envisages individual agents in which a certain adjustable parameter appears in each, and such that optimal controls (based on an agent-specific performance index) for each agent can be computed using the individual parameter and performance index.
Figure \ref{fig/multi_agent_problem} illustrates the arrangement. CT-MAOCS can be applied to multi-agent consensus problems where the shared information is a tunable parameter in the OC system of each agent. Parameter tuning needs to achieve a consensus while the optimal trajectory of each agent needs to satisfy a specific task specification under this consensus. An example treated in a later section is the synchronous multi-agent rendezvous problem \cite{lin2007rendezvous}, in which agents should determine their own optimal trajectories such that the rendezvous takes place at a certain specified time. The challenge in solving the problem of CT-MAOCS comes from two parts: first, each individual loss function $L_i$ is expressed using an explicit function both of the parameter $\boldsymbol{\theta}_i$ and the trajectory of the associated OC system, which makes the whole optimization problem at least a bi-level optimization; second, the optimization goal involves not just minimization of each agent's individual loss $L_i$ but the team-average loss, for which all tunable parameters need to be adjusted cooperatively. Main contribution of this paper is the development of a distributed framework to solve the problem of CT-MAOCS, which comes from a combination of a consensus-based distributed rule for multi-agent optimization in \cite{nedic2009distributed} and a gradient generator in \cite{jin2020pontryagin}.

\noindent \emph{Notations.}  Let $|\mathcal{N}|$ denote the cardinality of a set $\mathcal{N}$. Let $(\cdot)^\prime$ denote the Hermitian transpose. Let $||\cdot||$ denote the Euclidean norm. For a square matrix $A \in \mathbb{R}^{n \times n}$,   let $\text{Tr}(A)$ denote the trace of $A$. 
Let $\text{col}\{ \boldsymbol{v}_1, \cdots, \boldsymbol{v}_a \}$ denote a column stack of elements $\boldsymbol{v}_1, \cdots, \boldsymbol{v}_a $, which may be scalars, vectors or matrices,
i.e. $\text{col}\{ \boldsymbol{v}_1, \cdots, \boldsymbol{v}_a \} \triangleq {\matt{{\boldsymbol{v}_1}^{\prime} & \cdots & {\boldsymbol{v}_a}^{\prime}}}^{\prime}$.
Let $\frac{\partial\boldsymbol{g}_t}{\partial \boldsymbol{x}_t} \in \mathbb{R}^{n \times m}$ denote the Jacobian matrix of a function $\boldsymbol{g}: \mathbb{R}^n \mapsto \mathbb{R}^m$ with respect to $\boldsymbol{x} \in \mathbb{R}^n$  evaluated at $\boldsymbol{x}_t$, i.e., $\frac{\partial\boldsymbol{g}_t}{\partial \boldsymbol{x}_t}=\frac{\partial \boldsymbol{g}(\boldsymbol{x})}{\partial \boldsymbol{x}}\rvert_{\boldsymbol{x}=\boldsymbol{x}_t}$.


\section{Problem Formulation} \label{section:problem_formulation}



Consider a multi-agent system consisting of a number of $N$ agents labeled as $\mathcal{V}=\{1,\cdots,N\}$. Each agent $i$ can receive information from its neighbor set, which is denoted by $\mathcal{N}_i$. $\mathbb{G}_k = \{ \mathcal{V}, \mathcal{E}_k \}$ denotes the directed graph such that a directed edge from $j$ to $i$ is in $\mathcal{E}_k $ if and only if $j\in \mathcal{N}_i(k)$.

Suppose each agent $i$ is an optimal control system with a tunable parameter $\boldsymbol{\theta}_i \in \mathbb{R}^r$ denoted by $\boldsymbol{\mathcal{S}}_i(\boldsymbol{\theta}_i)$.  
The open-loop system dynamics of agent $i$, i.e. $\boldsymbol{\mathcal{S}}_i(\boldsymbol{\theta}_i)$,  are described by
$$\boldsymbol{x}_{i,t+1}=\boldsymbol{f}_i(\boldsymbol{x}_{i,t}, \boldsymbol{u}_{i,t}, \boldsymbol{\theta}_i),$$ where $t=0,\cdots, T$ denotes the discrete time index, $\boldsymbol{x}_{i,t} \in \mathbb{R}^{n}$ denotes agent-$i$'s state at time $t$,  $\boldsymbol{u}_{i,t} \in \mathbb{R}^{m}$ denotes agent-$i$'s optimal control input \footnote{ Optimal control inputs in this paper will be taken to be open-loop time functions rather than controls generated by a feedback law.}  at time $t$, and $\boldsymbol{f}_i:\mathbb{R}^{n}\times\mathbb{R}^{m}\times\mathbb{R}^r\mapsto\mathbb{R}^{n}$ denotes the nonlinear dynamics of agent-$i$, which is assumed to be twice-differentiable.
The open-loop control $\boldsymbol{u}_{i,t}$ could be replaced or determined by an optimal control determined in the following way. Associated with agent $i$ is an objective function denoted by
$$J_i = \sum\nolimits_{t=0}^{T-1}c_{i,t}(\boldsymbol{x}_{i,t},\boldsymbol{u}_{i,t}, \boldsymbol{\theta}_i)+\textcolor{black}{h_i(\boldsymbol{x}_{i,T},\boldsymbol{\theta}_i)},$$
where $c_{i,t}:\mathbb{R}^{n}\times\mathbb{R}^{m}\times\mathbb{R}^r\mapsto\mathbb{R}$ and $h_i:\mathbb{R}^{n}\times\mathbb{R}^r\mapsto\mathbb{R}$ denoting the running and final cost, respectively.
Then under a given initial condition $\boldsymbol{x}_{i,0}$, the optimal control for agent $i$ can be determined by
\begin{mini}|s|
	{\substack{\boldsymbol{x}_{i,1:T}, \\ \boldsymbol{u}_{i,0:T-1}}}{ J_i = \sum\nolimits_{t=0}^{T-1}c_{i,t}(\boldsymbol{x}_{i,t},\boldsymbol{u}_{i,t}, \boldsymbol{\theta}_i)+\textcolor{black}{h_i(\boldsymbol{x}_{i,T},\boldsymbol{\theta}_i)} \label{oc}}
	{}{}
	\addConstraint{ \boldsymbol{x}_{i,t+1} =\boldsymbol{f}_i(\boldsymbol{x}_{i,t},\boldsymbol{u}_{i,t}, \boldsymbol{\theta}_i)}
	\addConstraint{\forall t=0,\cdots,T-1 \  \text{with given } \boldsymbol{x}_{i,0},}
\end{mini}
Here $\boldsymbol{x}_{i,0:T} \triangleq \text{col}\{\boldsymbol{x}_{i,0},\cdots,\boldsymbol{x}_{i,T}\} \in \mathbb{R}^{n(T+1)}$ denotes all the states from time $t=0$ to $T$; similarly $\boldsymbol{u}_{i,0:T-1} \triangleq \text{col}\{\boldsymbol{u}_{i,0},\cdots,\boldsymbol{u}_{i,T-1}\} \in \mathbb{R}^{mT}$; the optimal control will be denoted by $\boldsymbol{u}^*_{i,0:T-1}$. Given a particular value of $\boldsymbol{\theta}_i$, the inputs $\boldsymbol{u}_{i,0:T-1}$ in (\ref{oc}) are designed to minimize the objective function $J_i$. For notational simplicity, let
$$ \boldsymbol{\xi}_i({{{\boldsymbol{\theta}}_i}}) \triangleq \text{col}\{\boldsymbol{x}_{i,0:T}, \boldsymbol{u}_{i,0:T-1} \} \in \mathbb{R}^{a}$$
denote the \emph{trajectory} of $\boldsymbol{\mathcal{S}}_i({\boldsymbol{\theta}}_i)$ given $\boldsymbol{\theta}_i$, where $a = (T+1)n+Tm$.
We assume, as is common, that any necessary smoothness and similar conditions for a well-defined unique solution to exist are fulfilled. Since this optimal trajectory depends on the parameter $\boldsymbol{\theta}_i$, $\boldsymbol{\xi}_i$ can also be viewed as a function of $\boldsymbol{\theta}_i$, i.e. $\boldsymbol{\xi}_i: \mathbb{R}^r \mapsto \mathbb{R}^a$.

Let $L_i(\boldsymbol{\xi}_i, \boldsymbol{\theta}_i)$ denote a scalar function, which is an additional `partial' performance index or loss function (independent of $J_i$) used as a contribution to a group or team-average performance index or loss function, and reflecting to agent $i$'s trajectory $\boldsymbol{\xi}_i({\boldsymbol{\theta}_i})$ as well as the associated performance index; thus ultimately, $L_i$ is just a function of $\boldsymbol{\theta}_i$ because $\boldsymbol{\xi}_i({\boldsymbol{\theta}_i})$ is in effect determined by the minimization of $J_i$. Correspondingly the global average $\frac{1}{N} \sum_{i=1}^{N}L_i(\boldsymbol{\xi}_i, \boldsymbol{\theta}_i)$ evaluates the performance of the whole multi-agent system. Note that the objective function $J_i$ reflects a task specification that is only related to agent-$i$, whereas the loss function $L_i$ indicates a new task specification which might be also related to other agents.

\textcolor{black}{
	The \textbf{problem of interest} is to develop an iterative rule for each agent $i$ to update $\boldsymbol{\theta}_i$ such that all the $\boldsymbol{\theta}_i$ reach a consensus at a common parameter $\boldsymbol{\theta}^*$, which minimizes the global average loss, i.e. 
	\begin{argmini}|s|
		{{\{\boldsymbol{\theta}_i\}}_{i=1}^N}{\frac{1}{N} \sum_{i=1}^{N} L_i(\boldsymbol{\xi}_i(\boldsymbol{\theta}_i), \boldsymbol{\theta}_i)}
		{\label{problem_interest}}{{\{\boldsymbol{\theta}^*\}}_{i=1}^N =}
		\addConstraint{\boldsymbol{\xi}_i \text{ obtained by (\ref{oc}) under } \boldsymbol{\theta}_i.}
\end{argmini}}
Note that the notation $\boldsymbol{\xi}_i$ is shorthand for the whole optimal trajectory (input and state) of agent-$i$ from $t=0$ to $t=T$.

\section{Main Results} \label{section:method}

The challenge in solving the cooperative tuning problem of multi-agent optimal control systems in (\ref{problem_interest}) comes from two parts: first, each $L_i$ here is a function of the trajectory of a dynamical system $\boldsymbol{\mathcal{S}}_i(\boldsymbol{\theta}_i)$, which makes the whole optimization problem at least a bi-level optimization; second, the optimization goal is not just the minimization of each system's own loss but the team-average loss, for which all tunable parameters need to be adjusted cooperatively.  Motivated by these two challenges, in this section we will develop a method to solve (\ref{problem_interest}) by a combination of consensus-based distributed optimization in \cite{nedic2009distributed} and a gradient generator in \cite{jin2020pontryagin}.

\subsection{Consensus-based Distributed Optimization} \label{subsec:optimization_update}

We first suppose the
gradient $\frac{d L_i(\boldsymbol{\xi}_i(\boldsymbol{\theta}_i), \boldsymbol{\theta}_i)}{d \boldsymbol{\theta}_i}$
is available for each agent $i$ (we shall explain in the next subsection how it can be obtained). Then the problem in (\ref{problem_interest}) becomes a standard consensus-based multi-agent optimization as follows:
\begin{mini}|s|
	{  \boldsymbol{\theta}_1,..., \boldsymbol{\theta}_N   }{ \frac{1}{N} \sum_{i=1}^{N} L_i(\boldsymbol{\xi}_i(\boldsymbol{\theta}_i), \boldsymbol{\theta}_i) \label{problem_consensus}}
	{}{}
	\addConstraint{  \boldsymbol{\theta}_1=\cdots = \boldsymbol{\theta}_N .}
\end{mini}

Let $k=0,1, \cdots$ denote the iteration index for adjustment of tunable parameters. Let $\boldsymbol{\theta}_i(k) \in \mathbb{R}^r$ denote agent $i$'s tunable parameter at iteration $k$.
At iteration $k$, the optimal control sequence $\boldsymbol{u}_{i,0:T-1}^*(k)$ for agent-$i$ is computed once based on \eqref{oc} with current parameter $\boldsymbol{\theta}_i(k)$. Then an updated value of $\boldsymbol{\theta}_i(k+1)$ will be computed using $\boldsymbol{\theta}_i(k)$ and the optimal control sequence $\boldsymbol{u}_{i,0:T-1}^*(k)$ together with other information from agent-$i$'s neighbors.

We in fact employ the following \emph{consensus-based gradient-descent update} proposed by \cite{nedic2009distributed, nedic2001incremental}: 
\begin{equation} \label{update_rule}
	\boldsymbol{\theta}_i(k+1) = \sum_{j \in \mathcal{N}_i(k)} w_{ij}(k) \boldsymbol{\theta}_j(k) - \eta(k) \frac{dL_i}{d\boldsymbol{\theta}_i(k) }.
\end{equation}
Here,  $$\frac{dL_i}{d\boldsymbol{\theta}_i(k)}=\frac{dL_i}{d \boldsymbol{\theta}_i }|_{ \boldsymbol{\theta}_i=\boldsymbol{\theta}_i(k)} $$ is the gradient of agent-$i$'s local loss $L_i$ with respect to $\boldsymbol{\theta}_i$ evaluated at $\boldsymbol{\theta}_i = \boldsymbol{\theta}_i(k)$; and the optimal control sequence $\boldsymbol{u}_{i,0:T-1}^*(k)$ is used in evaluating the gradient; further, $\eta(k) > 0$ is a diminishing step size for agent-$i$ such that
\begin{equation} \label{eq:step_size}
	\lim_{k \to \infty} \eta(k) = 0,\  \sum_{k=0}^{\infty} \eta(k) = \infty, \ \sum_{k=0}^{\infty} \eta(k)^2 < \infty,
\end{equation}
and $w_{ij}(k)$ are non-negative weights. 
Let $W(k) \in \mathbb{R}^{N \times N}$ be such that the $ij$-th entry is $w_{ij}(k)$ if $j\in \mathcal{N}_i(k)$ and 0, otherwise. As adopted in \cite{nedic2009distributed, nedic2001incremental}, we make the following assumption
\begin{assum} \label{Assum_consensus}
	$W(k)$ is doubly stochastic for all $k=0,1,\cdots$. There exists positive integers $\tau$ and $l$  such that the union of graphs $\mathbb{G}_{kl+1+\tau}, \mathbb{G}_{kl+2+\tau}, ..., \mathbb{G}_{(k+1)l+\tau}$ is strongly connected. 
\end{assum}
By analytical proofs and results in \cite{nedic2009distributed, nedic2001incremental}, one directly has the following lemma

\begin{lem} \label{lemma:multi_agent_optimization} \cite{nedic2001incremental}
	\textcolor{black}{
		Assume that the gradient $\frac{d L_i(\boldsymbol{\xi}_i(\boldsymbol{\theta}_i), \boldsymbol{\theta}_i)}{d \boldsymbol{\theta}_i}$ is known for each agent $i$, and each $L_i(\boldsymbol{\xi}_i(\boldsymbol{\theta}_i), \boldsymbol{\theta}_i)$ is convex in $\boldsymbol{\theta}_i$.
		Then the distributed update (\ref{update_rule}) with Assumption \ref{Assum_consensus} and step size $\eqref{eq:step_size}$ drives $\boldsymbol{\theta}_i(k) \to \boldsymbol{\theta}^*$ as $k \to \infty$ for all $i \in \mathcal{V}$ and $\boldsymbol{\theta}^*$ minimizes the global average, i.e.
		\begin{argmini}|s|
			{\boldsymbol{\theta}_1, \cdots, \boldsymbol{\theta}_N} {\frac{1}{N} \sum_{i=1}^{N} L_i(\boldsymbol{\xi}_i(\boldsymbol{\theta}_i), \boldsymbol{\theta}_i).}
			{\label{consensus}}{ {\{\boldsymbol{\theta}^*\}}_{i=1}^N =}
	\end{argmini}}
\end{lem}

Note that \cite{nedic2001incremental} proves Lemma \ref{lemma:multi_agent_optimization} when the step size satisfies \eqref{eq:step_size} and \cite{nedic2009distributed} proves Lemma \ref{lemma:multi_agent_optimization} when the step size is a positive constant.
{\color{black}
	Lemma \ref{lemma:multi_agent_optimization} hypothesises that  the gradient $\frac{d L_i(\boldsymbol{\xi}_i(\boldsymbol{\theta}_i), \boldsymbol{\theta}_i)}{d \boldsymbol{\theta}_i}$ is available to each agent $i$ for all iterations $k=0,1,\cdots$. From the chain rule, one has
	\begin{equation} \label{derivative_chain_rule}
		\frac{d L_i(\boldsymbol{\xi}_i, \boldsymbol{\theta}_i)}{d \boldsymbol{\theta}_i} = \frac{\partial L_i (\boldsymbol{\xi}_i, \boldsymbol{\theta}_i)}{\partial \boldsymbol{\xi}_i}  \frac{\partial \boldsymbol{\xi}_i ({\boldsymbol{\theta}}_i)}{\partial \boldsymbol{\theta}_i} + \frac{\partial L_i(\boldsymbol{\xi}_i, \boldsymbol{\theta}_i)}{\partial \boldsymbol{\theta}_i},
	\end{equation}
	where the partial derivatives $\frac{\partial L_i(\boldsymbol{\xi}_i, \boldsymbol{\theta}_i)}{\partial \boldsymbol{\xi}_i}$ and $\frac{\partial L_i(\boldsymbol{\xi}_i, \boldsymbol{\theta}_i)}{\partial \boldsymbol{\theta}_i}$ are known. The main challenge here comes from the fact that agent $i$ does not have an analytical relation between its system trajectory $\boldsymbol{\xi}_i$ and the parameter $\boldsymbol{\theta}_i$, and thus does not know $\frac{\partial \boldsymbol{\xi}_i ({\boldsymbol{\theta}}_i)}{\partial \boldsymbol{\theta}_i}$, i.e., the partial derivative of a trajectory $\boldsymbol{\xi}_i$ with respect to the parameter $\boldsymbol{\theta}_i$. In the following, we will borrow a result from \cite{jin2020pontryagin} to develop a \emph{gradient generator} which computes the exact value for $\frac{\partial \boldsymbol{\xi}_i(\boldsymbol{\theta}_i)}{\partial \boldsymbol{\theta}_i}$.
}

\subsection{Gradient Generator} \label{subsec:grad_generator}

This subsection introduces the gradient generator for computing $\frac{\partial \boldsymbol{\xi}_i( \boldsymbol{\theta}_i) }{\partial \boldsymbol{\theta}_i}$ at each iteration $k$. For simplicity of notation, we use $\boldsymbol{\theta}_i$ to denote $\boldsymbol{\theta}_i(k)$ in this section.
Given the optimal control (\ref{oc}), one has the following \emph{Hamiltonian} associated with $\boldsymbol{\mathcal{S}_i}(\boldsymbol{\theta}_i)$ for all $t=0, \cdots, T-1$,
\begin{equation}\label{Hamil}
	H_{i,t}=c_{i,t}(\boldsymbol{x}_{i,t},\boldsymbol{u}_{i,t},\boldsymbol{\theta}_i)+\boldsymbol f_i(\boldsymbol{x}_{i,t},\boldsymbol{u}_{i,t},\boldsymbol{\theta}_i)^\prime\boldsymbol{\lambda}_{i,t+1}.
\end{equation}
Here, $\boldsymbol{\lambda}_t \in \mathbb{R}^{n}, \ t=1,\cdots,T$ denotes the Lagrangian multiplier associated with the equality constraint which represents the dynamics $\boldsymbol{x}_{i,t+1}=\boldsymbol{f}_i(\boldsymbol{x}_{i,t},\boldsymbol{u}_{i,t}, \boldsymbol{\theta}_i)$.
By the definition of $\boldsymbol{\xi}_i$, we have
\begin{equation*}
	\frac{\partial \boldsymbol{\xi}_i(\boldsymbol{\theta}_i)}{\partial \boldsymbol{\theta}_i} = \matt{ \frac{\partial \boldsymbol{x}_{i,0:T}}{\partial \boldsymbol{\theta}_i} \\ \frac{\partial \boldsymbol{u}_{i,0:T-1}}{\partial \boldsymbol{\theta}_i} }.
\end{equation*}
Let
\begin{equation*}
	X_{i,t} \triangleq \frac{\partial \boldsymbol{x}_{i,t}}{\partial \boldsymbol{\theta}_i} \in \mathbb{R}^{n \times r}, \ U_{i, t} \triangleq \frac{\partial \boldsymbol{u}_{i,t}}{\partial \boldsymbol{\theta}_i} \in \mathbb{R}^{m \times r}.
\end{equation*}
Note that $X_{i,0} = \frac{\partial \boldsymbol{x}_{i,o}}{\partial \boldsymbol{\theta}_i} = \boldsymbol{0}$ because $\boldsymbol{x}_{i,0}$ in (\ref{oc}) is given. 



The tool for computing the gradient $\frac{\partial \boldsymbol{\xi}_i( \boldsymbol{\theta}_i) }{\partial \boldsymbol{\theta}_i}$ involves a linear quadratic control system  $\overline{\boldsymbol{\mathcal{S}}}_i(\boldsymbol{\theta}_i)$ given as follows:
\begin{mini}|s|
	{\substack{X_{i,1:T},\\U_{i,0:T-1}}} {\bar{J}_i = \text{Tr}\sum_{t=0}^{T-1}\Bigg(\frac{1}{2}\small
		\begin{bmatrix}
			{X}_{i,t} \\
			{U}_{i,t}
		\end{bmatrix}^\prime
		\bar{Q}_{i,t}
		\begin{bmatrix}
			{X}_{i,t} \\
			{U}_{i,t}
		\end{bmatrix}+
		\bar{R}_{i,t}^\prime
		\begin{bmatrix}
			{X}_{i,t} \\
			{U}_{i,t}
		\end{bmatrix}\Bigg) \label{auxiliary_lqr_system}}{}{}
	\breakObjective{\ \ \ \ \ \ + \text{Tr}\left(\frac{1}{2}{X}_{i,T}^\prime \, H_{i,T}^{xx} \,{X}_{i,T}+ (H_{i,T}^{x\theta})^\prime\,{X}_{i,T}\right)}
	\addConstraint{X_{i,t+1} = F_{i,t} X_{i,t} + G_{i,t} U_{i,t} + E_{i,t} \  \text{with} \  X_{i,0}=\boldsymbol{0}}
	\addConstraint{\bar{Q}_{i,t} = \begin{bmatrix}
			H_{i,t}^{xx} & H_{i,t}^{xu} \\
			H_{i,t}^{ux}& H_{i,t}^{uu}
		\end{bmatrix},
		\ \bar{R}_{i,t} = \begin{bmatrix}
			H_{i,t}^{x\theta} \\
			H_{i,t}^{u\theta}
		\end{bmatrix}.}
\end{mini}
The coefficients in (\ref{auxiliary_lqr_system}) are defined as follows:
\begin{flalign}
	&F_{i,t}=\dfrac{\partial \boldsymbol{f}_i}{\partial \boldsymbol{x}_{i,t}}, \ G_{i,t}=\dfrac{\partial \boldsymbol{f}_i}{\partial \boldsymbol{u}_{i,t}}, \ E_{i,t}=\dfrac{\partial \boldsymbol{f}_i}{\partial \boldsymbol{\theta}_i} \label{matFGEHx} & \\
	&H_{i,t}^{xx}=\dfrac{\partial^2 H_{i,t}}{\partial \boldsymbol{x}_{i,t}\partial \boldsymbol{x}_{i,t}}, \ H_{i,t}^{ux}=\dfrac{\partial^2 H_{i,t}}{\partial \boldsymbol{u}_{i,t} \partial \boldsymbol{x}_{i,t}}={(H_{i,t}^{xu})}^\prime, \label{matHux_and_Hxe} & \\
	&H_{i,t}^{uu}=\dfrac{\partial^2 H_{i,t}}{\partial \boldsymbol{u}_{i,t}\partial \boldsymbol{u}_{i,t}}, \ H_{i,t}^{x\theta}=\dfrac{\partial^2 H_{i,t}}{\partial \boldsymbol{x}_{i,t}\partial \boldsymbol{\theta}_i}, \  H_{i,t}^{u\theta}=\dfrac{\partial^2 H_{i,t}}{\partial \boldsymbol{u}_{i,t}\partial \boldsymbol{\theta}_i}, \label{matHuu_and_Hue} & \\
	&H_{i,T}^{xx}=\dfrac{\partial^2 h_i}{\partial \boldsymbol{x}_{i,T}\partial \boldsymbol{x}_{i,T}}, \  H_{i,T}^{x\theta}=\dfrac{\partial^2 h_i}{\partial \boldsymbol{x}_{i,T}\partial \boldsymbol{\theta}_i} \label{matHT}
\end{flalign}
which  are known based on the trajectory $\boldsymbol{\xi}_i(\boldsymbol{\theta}_i)$ and the trajectory of Lagrangian multipliers $\boldsymbol{\lambda}_{i,1:T}$. By the discrete-time Pontryagin's Maximum Principle \cite{jin2020pontryagin}, the Lagrangian multipliers $\boldsymbol{\lambda}_{i,1:T}$ can be obtained by iteratively computing (\ref{lagrangian_1}) and (\ref{lagrangian_2}) given $\boldsymbol{\xi}_i(\boldsymbol{\theta}_i)$:
\begin{flalign}
	&\boldsymbol{{\lambda}}_{i,T} = \frac{\partial h_i}{\partial\boldsymbol{{x}}_{i,T}}, \label{lagrangian_1} & \\
	&\boldsymbol{\lambda}_{i,t} \triangleq \dfrac{\partial H_{i,t}}{\partial \boldsymbol{{x}}_{i,t}} = \dfrac{\partial c_{i,t}}{\partial \boldsymbol{{x}}_{i,t}}+\dfrac{\partial \boldsymbol{f}_i^\prime}{\partial \boldsymbol{{x}}_{i,t}}\boldsymbol{{\lambda}}_{i,t+1}, \  t=T-1,\cdots, 1. \label{lagrangian_2}
\end{flalign}
In practice, many nonlinear optimization solvers, such as IPOPT \cite{wachter2006implementation}, can return the value of Lagrangian multipliers after a constrained nonlinear program is solved.

Note that $\overline{\boldsymbol{\mathcal{S}}}_i(\boldsymbol{\theta}_i)$ is of the linear quadratic regulator (LQR) form \cite{anderson1990optimal} and the system dynamics and control objective in $\overline{\boldsymbol{\mathcal{S}}}_i(\boldsymbol{\theta}_i)$ are purely determined by the trajectory $\boldsymbol{\xi}_i(\boldsymbol{\theta}_i)$ from $\boldsymbol{\mathcal{S}}_i(\boldsymbol{\theta}_i)$.We also call $\overline{\boldsymbol{\mathcal{S}}}_i(\boldsymbol{\theta}_i)$ the \emph{gradient generator} because of the following lemma:


\begin{lem} \label{lemma:stationary} \cite[Lemma~5.1]{jin2020pontryagin} Let $\{ X_{i,0:T}^{*}, U_{i,0:T-1}^{*} \}$ be a stationary solution to (\ref{auxiliary_lqr_system}). Then
	\smallskip
	
	\begin{equation} \label{stationary_eq}
		\matt{ X_{i,0:T}^* \\ U_{i,0:T-1}^* } = \matt{ \frac{\partial \boldsymbol{x}_{i,0:T}}{\partial \boldsymbol{\theta}_i} \\ \frac{\partial \boldsymbol{u}_{i,0:T-1}}{\partial \boldsymbol{\theta}_i} } = \frac{\partial \boldsymbol{\xi}_i(\boldsymbol{\theta}_i)}{\partial \boldsymbol{\theta}_i}.
	\end{equation}
\end{lem}
By stationary solution we mean that $\{ X_{i,0:T}^{*}, U_{i,0:T-1}^{*} \}$ might be a saddle point or a minimum to (\ref{auxiliary_lqr_system}).
However, as long as $\{ X_{i,0:T}^{*}, U_{i,0:T-1}^{*} \}$ is a stationary solution to (\ref{auxiliary_lqr_system}), i.e. the gradients of (\ref{auxiliary_lqr_system}) are zeros, $\{X_{0:T}^*, U_{0:T-1}^*\}$ is exactly $\frac{\partial \boldsymbol{\xi}_i(\boldsymbol{\theta}_i)}{\partial \boldsymbol{\theta}_i}$. Since $\overline{\boldsymbol{\mathcal{S}}}_i(\boldsymbol{\theta}_i)$ is a linear quadratic control system, we can compute $\{X_{0:T}^*, U_{0:T-1}^*\}$ by the following lemma:

\begin{lem} \label{lemma:compute} \cite[Lemma~5.2]{jin2020pontryagin} $\{X_{0:T}^*, U_{0:T-1}^*\}$ can be obtained by the following recursions for $t=T-1, \cdots, 0$
	\begin{equation} \label{recursion_eq}
		\begin{aligned}
			P_{i,t} &= Q_{i,t} + A_{i,t}^{\prime}{(I+P_{i,t+1}R_{i,t})}^{-1}P_{i,t+1}A_{i,t}, \\
			W_{i,t} &= A_{i,t}^{\prime}{(I+P_{i,t+1}R_{i,t})}^{-1}(W_{i,t+1}+P_{i,t+1}M_{i,t}) + N_{i,t},
		\end{aligned}
	\end{equation}
	where $P_{i,T}=H_{i,T}^{xx}$, $W_{i,T}=H_{i,T}^{x\theta}$; $I$ is identity matrix; $A_{i,t} \triangleq F_{i,t}-G_{i,t}{(H_{i,t}^{uu})}^{-1}H_{i,t}^{ux}$, $R_{i,t} \triangleq G_{i,t}{(H_{i,t}^{uu})}^{-1}G_{i,t}^{\prime}$, $M_{i,t} \triangleq E_{i,t}-G_{i,t}{(H_{i,t}^{uu})}^{-1}H_{i,t}^{u\theta}$, $Q_{i,t} \triangleq H_{i,t}^{xx}-H_{i,t}^{xu}{(H_{i,t}^{uu})}^{-1}H_{i,t}^{ux}$, $N_{i,t} \triangleq H_{i,t}^{x\theta}-H_{i,t}^{xu}{(H_{i,t}^{uu})}^{-1}H_{i,t}^{u\theta}$. Further, $\{X_{0:T}^*, U_{0:T-1}^*\}$ can be computed by iteratively computing the following equations from $t=0$ to $T-1$ with $X_{i,0} = \boldsymbol{0}$:
	\begin{equation}  \label{recursion_solution:1}
		\begin{aligned}
			U_{i,t} = &- {(H_{i,t}^{uu})}^{-1}\Big( H_{i,t}^{ux}X_{i,t}+H_{i,t}^{u\theta} + G_{i,t}^{\prime}(I+P_{i,t+1} \cdot \\
				& {R_{i,t})}^{-1} \cdot (P_{i,t+1}A_{i,t}X_{i,t}+P_{i,t+1}M_{i,t}+W_{i,t+1}) \Big),
		\end{aligned}
	\end{equation}
	\begin{equation} \label{recursion_solution:2}
		X_{i,t+1}=F_{i,t}X_{i,t}+G_{i,t}U_{i,t}+E_{i,t}.
	\end{equation}
\end{lem}
\begin{rem}
	$H_{i,t}^{uu}$ in (\ref{recursion_solution:1}) for all $t = 0, \cdots, T-1$ is invertible if the second-order optimality sufficient condition of (\ref{oc}) is satisfied (as proved in Lemma 1 and Theorem 1 in \cite{jin2021safe}).
	See \cite[Lemma~A.2]{jin2021safe} for further details about the second-order sufficient condition.
	This is because when the condition holds, the Hessian matrix of the Hamiltonian in (\ref{Hamil}), $\matt{ H_{i,t}^{xx} & H_{i,t}^{xu} \\ H_{i,t}^{ux} & H_{i,t}^{uu} }$,  is a positive definite matrix for all $t = 0, \cdots, T-1$. This indicates that $H_{i,t}^{uu}$ is a positive definite matrix for all $t = 0, \cdots, T-1$, i.e. $H_{i,t}^{uu}$ in (\ref{recursion_solution:1}) is invertible. In this case, the stationary solution becomes a globally unique solution. If the second-order optimality sufficient condition does not hold, then one cannot use the recursions in Lemma \ref{lemma:compute} to compute a stationary solution to (\ref{auxiliary_lqr_system}). Nevertheless one can compute a stationary solution with a gradient descent-based method \cite{boyd2004convex}.
\end{rem}

\subsection{The Framework for Cooperative Tuning of Multi-Agent Optimal Control}
\begin{figure}[h]
	\centering
	\includegraphics[width=0.43\textwidth]{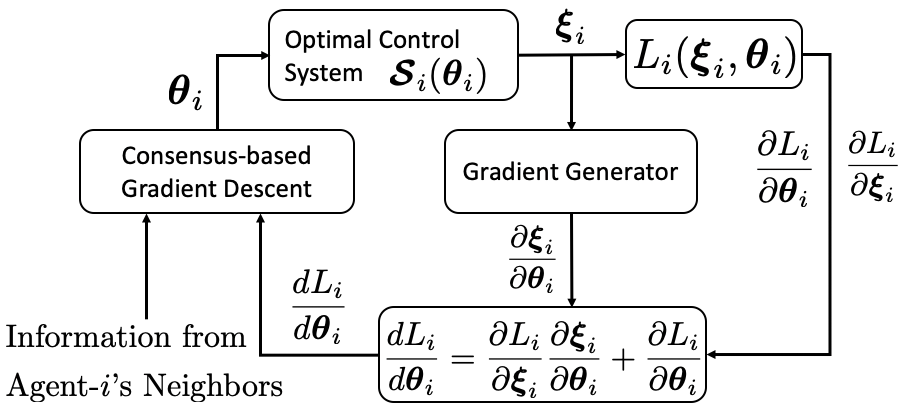}
	\caption{The framework for cooperative tuning}
	\label{fig/framework}
\end{figure}

To sum up, we employ the following framework for cooperative tuning of Multi-Agent Optimal Control, i.e. to solve the problem in (\ref{problem_interest}). This framework is based on a combination of the consensus-based gradient descent algorithm in (\ref{update_rule}) and the gradient generator in (\ref{auxiliary_lqr_system}), as shown in Fig. \ref{fig/framework}.

By Lemma \ref{lemma:multi_agent_optimization}, Lemma \ref{lemma:stationary} and Lemma \ref{lemma:compute}, one has the following main result.
\begin{thm} \label{theorem:1} Suppose that Assumption \ref{Assum_consensus} holds. The distributed update (\ref{update_rule}) is utilized for (\ref{problem_interest}), where $\frac{d L_i(\boldsymbol{\xi}_i, \boldsymbol{\theta}_i)}{d \boldsymbol{\theta}_i}$ is computed by the chain rule in (\ref{derivative_chain_rule}) and $\frac{\partial \boldsymbol{\xi}_i(\boldsymbol{\theta}_i)}{\partial \boldsymbol{\theta}_i}$ is obtained by the gradient generator (\ref{auxiliary_lqr_system}). One has all $\boldsymbol{\theta}_i(k) \to \boldsymbol{\theta}^*$ as $k \to \infty$ for all $i \in \mathcal{V}$ where $\boldsymbol{\theta}^*$ solves the problem in (\ref{problem_interest}).
\end{thm}


\subsection{Constraints in Optimal Control}

In the optimal control problem \eqref{oc}, one can add inequality constraints that represent safety constraints. With the interior-point method \cite{fiacco1990nonlinear}, one can define a logarithmic barrier function for each inequality constraint and a barrier parameter. Then the constrained optimization problem can be written as an unconstrained one, where the new objective function is the original one minus the summation of all the barrier functions. Hence, one can formulate a similar gradient generator for this new optimal control problem. The Hamiltonian associated with this new problem also includes the inequality constraint. See \cite{jin2021safe} for details.

\section{Simulation} \label{section:example}

This section applies the proposed cooperative tuning into a synchronous multi-agent rendezvous problem \cite{lin2007rendezvous}. Suppose there are $N$ mobile robots (or agents) and each agent should determine an optimal trajectory based on its optimal control. The rendezvous should take place at a certain specified time (i.e. the end of the trajectory), and the desired rendezvous location for each agent is unspecified, which is initialized randomly and viewed as a tunable parameter in the OC system of each agent.
Given a particular value of the tunable parameter for each agent, the determination of the optimal trajectory is made independently of the other agents.

At first iteration ($k=0$), each agent determines an optimal trajectory under a initial parameter $\boldsymbol{\theta}_i(k=0)$. Then for each iteration, agents share and update their parameters and cooperatively minimize a global loss function by individually minimizing their own local loss function. All the agents should eventually achieve a consensus on the parameter and hence rendezvous at a single unspecified location.

Agent-$i$'s dynamics are modeled by the following unicycle model \cite[Chapter~13]{lavalle2006planning}:
\begin{equation}\label{cov_model}
	\dot{\boldsymbol{x}}_i = \begin{bmatrix}
		\dot{p}_{x,i} \\
		\dot{p}_{y,i} \\
		\dot{\psi}_i
	\end{bmatrix} = \boldsymbol{f}_c(\boldsymbol{x}_i, \boldsymbol{u}_i) = \begin{bmatrix}
		u_{v,i} \cdot \text{cos}(\psi_i) \\
		u_{v,i} \cdot \text{sin}(\psi_i) \\
		u_{\omega, i}
	\end{bmatrix},
\end{equation}
where $\boldsymbol{x}_i \in \mathbb{R}^3$ is agent-$i$'s state, $\boldsymbol{u}_i = \text{col}\{ u_{v,i}, \ u_{\omega,i} \} \in \mathbb{R}^2$ is agent-$i$'s control input, $p_{x,i} \in \mathbb{R}$ and $p_{y,i} \in \mathbb{R}$ are position coordinates, $\psi_i \in \mathbb{R}$ is the heading angle, $u_{v,i}$ is the velocity input, and $u_{\omega,i}$ is the angular velocity input. Define 
\begin{equation}
	p: \boldsymbol{x}_i \in \mathbb{R}^3 \mapsto \boldsymbol{p}_i \in \mathbb{R}^2
\end{equation}
as the static mapping from agent-$i$'s state to its position $\boldsymbol{p}_i = \text{col}\{p_{x,i}, \ p_{y,i}\} \in \mathbb{R}^2$.

The optimal control for agent-$i$ is written as
\begin{mini}|s|
	{\substack{\boldsymbol{x}_{i,1:T}, \\ \boldsymbol{u}_{i,0:T-1}}}{J_i(\boldsymbol{x}_{i,0:T}, \boldsymbol{u}_{i,0:T-1}, \boldsymbol{\theta}_i) \label{multi_agent_rendezous}}
	{}{}
	\addConstraint{ \boldsymbol{x}_{i,t+1} = \boldsymbol{x}_{i,t} +  {\Delta} \cdot \boldsymbol{f}_c(\boldsymbol{x}_{i,t},\boldsymbol{u}_{i,t}) }
	\addConstraint{\forall t=0,\cdots,T-1 \  \text{with given } \boldsymbol{x}_{i,0},}
\end{mini}
where $\boldsymbol{\theta}_i \in \mathbb{R}^2$ is the tunable parameter for agent-$i$, $\Delta > 0$ is a constant arising in the discrete time Euler approximation of the differential equation (\ref{cov_model}), and the objective function $J_i(\boldsymbol{x}_{i,0:T}, \boldsymbol{u}_{i,0:T-1}, \boldsymbol{\theta}_i)$ is defined by
\begin{equation} \label{simulation_cost_function}
	J_i= \sum_{t=0}^{T-1} \Big[ 2||\boldsymbol{p}(\boldsymbol{x}_{i,t}) - \boldsymbol{\theta}_i ||^2 + ||\boldsymbol{u}_{i,t}||^2 \Big] 
	+ 5||\boldsymbol{p}(\boldsymbol{x}_{i,T}) - \boldsymbol{\theta}_i||^2.
\end{equation}
The local loss function for agent-$i$ is defined by
\begin{equation} \label{example_local_loss}
	L_i(\boldsymbol{\xi}_{i}, \boldsymbol{\theta}_i) = 100||\boldsymbol{p}(\boldsymbol{x}_{i,T}) - \boldsymbol{\theta}_i||^2
\end{equation}
where $\boldsymbol{\xi}_i \triangleq \text{col}\{\boldsymbol{x}_{i,0:T}, \boldsymbol{u}_{i,0:T-1} \} \in \mathbb{R}^{5T+3}$, and $\boldsymbol{x}_{i,T}$ is the $(T+1)$-th component of $\boldsymbol{\xi}_i$. And the global loss function is $\frac{1}{N} \sum_{i=1}^N L_i$. Note that the weighting coefficients in (\ref{simulation_cost_function}) and (\ref{example_local_loss}) are essentially arbitrary.

Section \ref{section:problem_formulation} mentions the difference between an objective function $J_i$ and a local loss function $L_i$ in general.
In this specific example, $J_i$ (through its inclusion of the term $||\boldsymbol{u}_{i,t}||^2$) indicates that the trajectory of each agent should seek over the whole trajectory to become as close as possible to the desired rendezvous location while maintaining small energy consumption, whereas $L_i$ indicates that the end of the trajectory should be as close as possible to the desired rendezvous location, and nothing more than that, since energy use and proximity to the rendezvous point before the end-time are irrelevant to the global objective.

\begin{figure*}[h]
	\centering
	\begin{subfigure}{.329\textwidth}
		\centering
		\includegraphics[width=\linewidth]{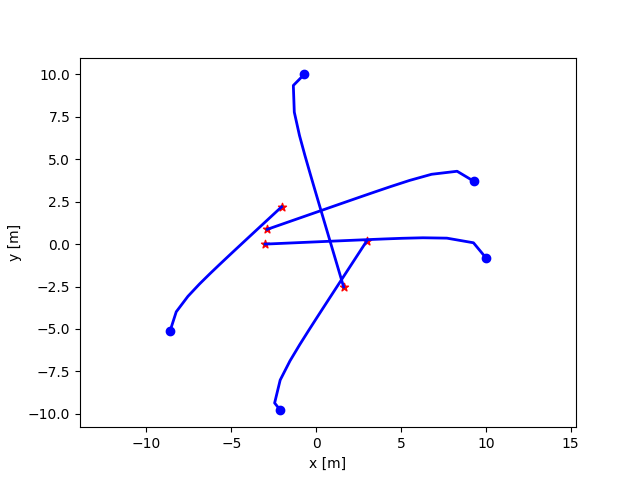}
		\caption{The trajectory before iteration}
		\label{fig:traj_before}
	\end{subfigure}
	\hfill
	\begin{subfigure}{.329\textwidth}
		\centering
		\includegraphics[width=\linewidth]{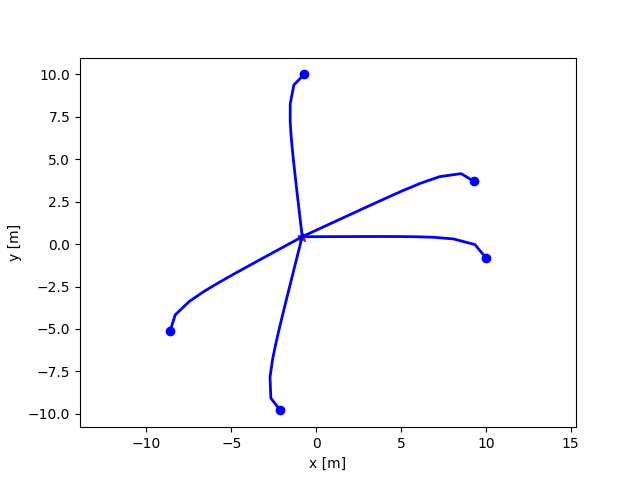}
		\caption{The trajectory after 30 iterations}
		\label{fig:traj_after}
	\end{subfigure}
	\hfill
	\begin{subfigure}{.329\textwidth}
		\centering
		\includegraphics[width=\linewidth]{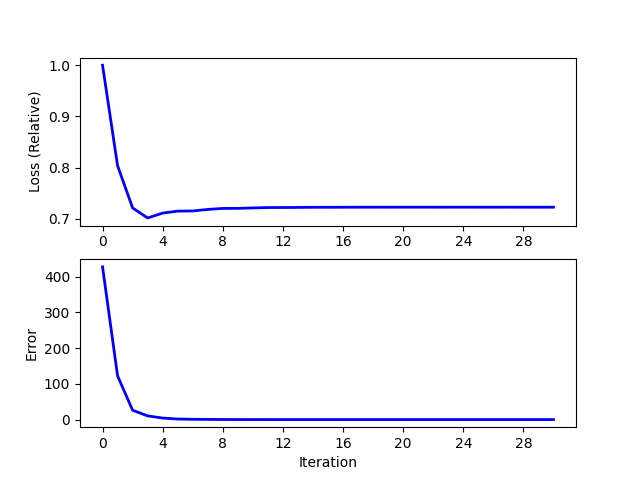}
		\caption{The loss and parameter error}
		\label{fig:loss_traj_}
	\end{subfigure}
	\caption{ The simulation result for a multi-agent rendezvous problem given a periodic graph with 5 agents. The blue dots are the initial positions. The red stars are the desired terminal positions $\boldsymbol{\theta}_i$. The lines in blue are the optimal trajectory generated by the optimal controls given $\boldsymbol{\theta}_i$. The top plot in (c) is relative loss over iterations, i.e., current loss divided by the initial loss. The bottom plot in (c) is total error of parameter $\boldsymbol{\theta}_i$ among all agents over iterations, i.e., $\sum_{i=1}^N \sum_{j=1}^N ||\boldsymbol{\theta}_i - \boldsymbol{\theta}_j||^2$. }
	\label{fig:loss_traj}
\end{figure*}

\subsection{Simulation Result} \label{subsec:simulation_result}

\begin{figure}[h]
	\centering
	\includegraphics[width=0.43\textwidth]{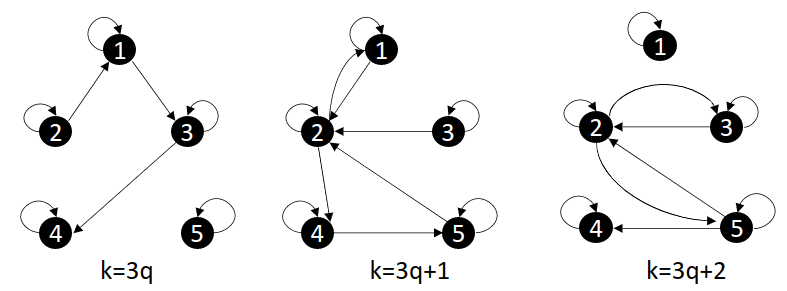}
	\caption{Periodic time variant graph $\mathbb{G}_k$, $q = 0, 1, 2, \cdots$}
	\label{fig:graph}
\end{figure}

The other parameters used for the following simulation are: $N=5$, $T=60$, $\Delta=0.1$s, $\eta(k) = 0.1 \ \forall k \geq 0$. A periodic time variant graph $\mathbb{G}_k$ is defined in Fig. \ref{fig:graph}. The weight matrix $W(k)$ is defined by Metropolis weights \cite{xiao2005scheme}. The initial state $\boldsymbol{x}_{i,0}$ and parameter $\boldsymbol{\theta}_i(0)$ are generated randomly.

As shown in Fig. \ref{fig:loss_traj}(a) and \ref{fig:loss_traj}(b), the tunable parameters $\boldsymbol{\theta}_i$ are initialized as different positions at first iteration. As the iteration $k$ increases, the $\boldsymbol{\theta}_i(k)$ converge to a common point, resulting in multiple agents rendezvousing with each other. In Fig. \ref{fig:loss_traj}(c), the loss is decreasing when the parameter error $\sum_{i=1}^N \sum_{j=1}^N ||\boldsymbol{\theta}_i - \boldsymbol{\theta}_j||^2$ is decreasing significantly, and finally both the loss and the parameter error converge.

\section{Conclusion} \label{section:discussion}

This paper has developed a framework based on a combination of consensus-based distributed optimization and gradient generator, which solves the problem of cooperative tuning of multi-agent optimal control system.
Future work include development of a gradient estimator based on trajectory segments of optimal control systems, extension of the result to optimal control systems with infinite time horizon and employment of other gradient-descent algorithms, such as Nesterov’s Accelerated Gradient \cite{sutskever2013importance}.

	\bibliographystyle{ieeetr}
	
	\bibliography{reference}

\end{document}